\pdfoutput=1
\documentclass[a4paper,twoside]{article}
\newcommand{\affil}[1]{$^{\rm #1}$}
\usepackage[authoryear]{natbib}
\bibpunct{(}{)}{;}{a}{}{,}
\usepackage{graphicx}
\date{} 
\title{\large\bf\flushleft Searching for Faint Planetary Nebulae Using the Digital Sky Survey}
\author{\parbox{\textwidth}{\flushleft
\vspace{-0.5cm}
{\it George H. Jacoby\affil{A,H}, Matthias Kronberger\affil{B}, Dana Patchick\affil{B}, 
Philipp Teutsch\affil{B,C}, Jaakko Saloranta\affil{B}, Michael Howell\affil{B}, Richard Crisp\affil{B},
Dave Riddle\affil{B}, Agn\'es Acker\affil{D}, David J. Frew\affil{E,F}, and Quentin Parker\affil{E,G}} \\
\vspace{0.4cm}
{\small \affil{A}\,WIYN Observatory, 950 N Cherry Ave, Tucson, AZ, 85719, USA}\\
{\small \affil{B}\,Deepskyhunters Collaboration}\\
{\small \affil{C}\,Institut f\"ur Astrophysik, Leopold-Franzens-Universit\"at Innsbruck, Austria}\\
{\small \affil{D}\,Observatoire Astronomique, Universit\'e de Strasbourg, 
67000 Strasbourg, France}\\
{\small \affil{E}\,Department of Physics, Macquarie University, Sydney NSW 2109, Australia}\\
{\small \affil{F}\,Perth Observatory, Bickley, WA 6076, Australia}\\
{\small \affil{G}\,Anglo-Australian Observatory, Epping, NSW1710, Australia}\\
{\small \affil{H}\,Email: jacoby@noao.edu}}}
\begin{document}
\twocolumn[
\begin{changemargin}{.8cm}{.5cm}
\begin{minipage}{.9\textwidth}
\vspace{-1cm}
\maketitle
%
%
\small{\bf Abstract:}
Recent H$\alpha$ surveys such as SHS and IPHAS have improved the completeness of the Galactic planetary nebula (PN) census. We now know of $\sim$3~000 PNe in the Galaxy, but this is far short of most estimates, typically $\sim25~000$ or more for the total population. The size of the Galactic PN population
is required to derive an accurate estimate of the chemical enrichment rates of nitrogen, carbon, and helium. In addition, a high PN count ($>$20~000) is strong evidence that most 1-8 M$_\odot$ main sequence stars will go through a PN phase, while a low count ($<$10~000) argues that special conditions (e.g., a close binary interaction) are required to form a PN. We describe a technique for finding hundreds more PNe using the existing data collections of the digital sky surveys, thereby improving the census of Galactic PNe.


\medskip
\medskip
\end{minipage}
\end{changemargin}
]
\small


\section{Introduction}

Counting the number of planetary nebulae (PNe) in the Galaxy is not just a bookkeeping exercise. The PN phase enriches the ISM with nitrogen, carbon, helium, and dust, important components in the formation of future generations of stars. Also, PNe may be progenitors of Type Ia supernovae, which return prodigious amounts of iron to
the ISM; a superb candidate is the pair of central stars in PN G135.9+55.9 (Tovmassian et al.\ 2007). \cite{miszalski09} find several similar short-period central star binaries. Thus, we need to know the number of PNe in the Galaxy in order to develop accurate models of the chemical enrichment rates since the initial burst of Type II supernovae when the Galaxy was young.

More directly, the number count of PNe in the Galaxy depends strongly on whether single stars like the Sun can form a PN, or if more unusual circumstances are required, such as the presence of a close binary companion. That is, the true number of PNe depends on how stars evolve and how PNe form, and so, a simple count provides a strong test of one of the predictions of stellar evolution -- that all low mass stars between 1--8 M$_{\odot}$ will go through a PN phase. If this is true, there will be a much larger number of PNe in the Galaxy than if a special condition is required. The models of \cite{moe06}, for example, predict that there are $46~000\pm22~000$ for the general case, but only $\sim6~600$ (De Marco \& Moe 2005) if close binaries (e.g., common envelope phase) are required. \cite{miszalski09} confirm earlier estimates that the binary fraction of PN central stars is only 10-20\% of all PNe, and thus, binarity is not likely to be a major factor in the formation process. If true, there should 
be many PNe waiting to be found in the Galaxy. 

The goal of this study is to demonstrate another method toward improving the Galactic census of PNe. We summarize the census problem in section 2, describe our technique in section 3, and report on the first results of our survey in section 4. Section 5 describes some implications of the results.


\section{Overview of the Problem}

It is almost impossible to count all the PNe in the Galaxy directly. The most serious obstacle is dust extinction in the Galactic plane where most PNe reside, rendering a large fraction of the Galaxy unobservable at optical wavelengths where PNe are most luminous. The identification problem is complicated further by confusion with HII regions, supernova remnants, distant galaxies, nova shells, and symbiotic nebulae.

A popular method for estimating the total number of PNe in the Galaxy is based on the identification of a complete sample of PNe within a local volume near the Sun and extrapolating that PN density (usually relative to either mass or luminosity) to the entire Milky Way. To do so, though, requires knowing distances to the local sample. PN distances are notoriously inaccurate at the 20-100\% level, leading to errors in the estimates of about 2--10$\times$. Consequently, this approach yields total counts that have a wide spread in values -- 
from 13~000 (Frew 2008) to 140~000 (Ishida \& Weinberger 1987), but most estimates derived this way  are close to $\sim$25~000 (see Frew 2008 for an excellent discussion of the various methods and results).

Dust extinction is a far less serious problem when counting PNe in nearby external galaxies, especially early-types (e.g, ellipticals) that have little or no dust, or nearby galaxies like the LMC where we can see all the PNe (Reid \& Parker 2006) down to extremely faint luminosities. That is, we can derive a reasonably accurate count of the PNe in galaxies outside our own, at least down to some limiting magnitude. We then extrapolate that count to the faintest magnitudes using the PN luminosity function, or PNLF (Ciardullo et al.\ 1989), as a guide. In comparison to the method of extrapolating the local volume to the entire Galaxy, the extragalactic method requires a much smaller stretch by a factor of several tens, thereby significantly reducing the risk in extrapolation. 

Once we have the external galaxy count, one can derive a PN production rate -- the number of PNe found in the galaxy, $N(PN)$, divided by the galaxy's bolometric luminosity, $L$. \cite{ciardullo05} find this parameter to be constant to within a factor of two for bright galaxies like the Milky Way. When multiplied by the estimated luminosity for our Galaxy, the $N(PN)/L$ predicts a total PN count in the Milky Way of $\sim$8~600 (Jacoby 1980; Peimbert 1990). While the counts based on external galaxies are less direct and do not actually identify the PN in our Galaxy, they provide a less compromised approach for comparison with more direct methods, as summarized in Table 1.

Another way to avoid the errors in PN distances while minimizing the effects of dust is to calculate the density of PNe in clear regions toward the Galactic bulge, whose distance is reasonably well determined at 8.4$\pm0.6$ kpc (Reid et al.\ 2009). Using the ratio of the bulge to total mass of the Galaxy, one can extrapolate for the total population. \cite{acker92a} and \cite{peyaud05} derive values of 17~000 and 35~000, respectively, using this technique.

The actual count of known PNe is in strong contrast to any PN population estimates. Over the last 85 years of PN surveys, the community has verified the existence of only $\sim$3~000, despite the use of a wide variety of techniques (e.g., radio, IRAS colors, POSS plates, optical and near-IR imaging). Adding to previous PN collections (Acker et al.\ 1992b; Acker et al.\ 1996; Kohoutek 2001), \cite{parker06} and \cite{miszalski08} used the southern H$\alpha$ survey (SHS) by \cite{parker05} to add 1~238 PNe (MASH-I and MASH-II), and using the survey data from \cite{drew05} for a smaller region in the north (IPHAS), \cite{viironen09a} added another 289 PNe (see also Viironen et al.\ 2009b). These two surveys doubled the previously known PN census -- a significant increment, but far short of the predictions. 

Table 1 summarizes several representative Galactic census predictions. Caution must be exercised in comparing these values since assumptions on maximum observable lifetimes or diameters are required in some cases. Still, we are missing half the PNe relative to the smallest estimate, which is based on the hypothesis that PNe form only from a common envelope interaction. Relative to the estimates from external galaxies, the lowest observationally based values, we have yet to identify two-thirds of the Galactic PNe. Comparisons with the other estimates are far more discrepant, exceeding 90\% incompleteness in some cases. These large differences demonstrate that something is seriously wrong. Among the possible ways to resolve the disparity in the PN census estimates, we have:

\begin{enumerate}
\item{Dust obscuration is hiding 50\% to $>$90\% of the PNe in the Galaxy (Jacoby \& Van de Steene 2004; Parker et al. 2006; Miszalski et al.\ 2008). This is the leading explanation.}

\item{Some biases act to prevent finding some classes of PNe (e.g., very compact, very low surface brightness).}

\item{Some stellar evolution process exists that prevents PNe from surviving as long as we expect, typically 25~000--50~000 yrs (e.g., disruptions due to interactions with the ISM -- Wareing et al. 2006).}

\item{Extrapolation from the local volume is unreliable because distances are incorrect (Napiwotzki 2001). Depending on the extrapolation technique, distance enters as the $3^{rd}$ or $4^{th}$ power. A simple volume extrapolation (e.g., Cahn \& Wyatt 1976) is affected as the $3^{rd}$ power of the distance; if one estimates the nebula age from the radius to form a uniformly limited sample in lifetime, though, then the extrapolation is sensitive to the $4^{th}$ power of the distance. For example, a 20\% error in distance yields an error in the estimated count of 2$\times$. A more typical error in distance of 50\% yields a count error of $\sim$4$\times$. To explain the discrepancy between the current census and the estimates for the total population, the average distances must always be too small, leading to an overestimated density. It is unclear, though, why the distance bias always falls on the short side.}

\item{There remains a very large area of the sky beyond Galactic latitudes of 10$^\circ$ that has not been surveyed as intensely as the lower latitudes.}

\item{There really are many fewer PNe in the Galaxy than the extrapolations suggest.}
\end{enumerate}

\begin{table}[h]
\begin{center}
\caption{Representative Estimates$^a$ of the Number of PNe in the Milky Way}\label{table1}
\begin{tabular}{lr}
\\
\hline Method & Count \\
\hline & \\
Actual Number Known$^b$     &  3~000  \\
\\
Population Synthesis I$^c$  & 46~000  \\
Population Synthesis II$^d$ &  6~600  \\
External galaxies$^e$       &  8~600  \\
Local Mass density$^f$      &140~000  \\
Local Luminosity density$^g$& 30~000  \\
Local space density$^h$     & 13~000  \\
\hline & 
\end{tabular}
\end{center}

$^a$Most of the estimates in Table 1 assume completeness for PNe having radii $<$0.9 pc. The actual number of known PNe includes nebulae of all radii. \\
$^b$ \cite{parker06}. \\
$^c$Assumes all stars (1-8 M$_{\odot}$) make a PN (Moe \& De Marco 2006). \\
$^d$Assumes only close binaries (common envelope) make a PN (De Marco \& Moe 2005). \\
$^e$Observationally based estimate, using PN luminosity function to extrapolate; average of values from \cite{peimbert90} and \cite{jacoby80} who report 7~200 and 10~000, respectively. \\
$^f$Observationally based estimate using the local PN surface density per unit Galactic mass (Ishida \& Weinberger 1987). \\
$^g$Observationally based estimate using the local PN density per unit Galactic luminosity (Phillips 2002). \\
$^h$Observationally based estimate, using local density to extrapolate to entire Galaxy volume (Frew 2008). 

\end{table}

By improving the census of PNe in the Galaxy, we improve our understanding of stellar evolution, the PN formation process, and/or the structure of our Galaxy. This is the motivation to improve the completeness of the Galactic PN surveys, which generally are highly heterogeneous in their depth. We have devised a search technique that uses existing all-sky data in the form of the digitized sky surveys. 


\section{The Survey Method}

A group of (mostly) amateur astronomers known as the Deepskyhunters (Kronberger et al.\ 2006) have been scanning the digital sky surveys (DSS), such as the POSS-I and POSS-II surveys, for open clusters for several years, and have been successful in finding numerous previously unknown clusters. In addition to finding cluster candidates, \cite{kronberger06} also identified 36 PN candidates. They have continued the search for both classes of objects and have been especially prolific with the PN search. The digitized photographic material extends very faint in surface brightness when examined by a trained eye for PN candidates. Their process is outlined as follows:

\begin{enumerate}
\item{Visually examine the digital sky survey data sets (e.g., POSS-II red and blue) for extended objects that may be PNe in areas of the sky that are generally complementary to the SHS and IPHAS surveys. Typically, our survey regions have targeted the Galactic plane beyond latitudes $>|5^\circ|$ north of declination $+2^\circ$ and latitudes $>|10^\circ|$ south of declination $+2^\circ$, but the search sometimes exceeds these limits.}
\item{Flag anything that looks like a PN.}
\item{Check that the object does not appear in 2MASS, although a few brighter candidates can be seen in the 2MASS J band image.}
\item{Check that the object does not appear in POSS-II IR image, which would suggest a continuum source (Cappellaro et al.\ 1990).}
\item{Check that the object is not already cataloged (e.g., in SIMBAD or in the PN catalogs of Acker et al.\ 1992b and Kohoutek 2001).}
\item{Cross check the candidate on other POSS-I and POSS-II images to guard against plate defects.}
\item{The candidate has a higher probability of being a real PN if it has a faint blue star near its center.}
\item{Send candidate lists to professional colleague for imaging in H$\alpha$ (mostly with the WIYN 3.5-m or OHP 1.2-m telescopes). WIYN observations were 300-s exposures using the OPTIC camera (Howell et al.\ 2003). OHP observations used 120-s to 300-s exposures.}
\item{Evaluate the H$\alpha$ morphology to assess likelihood of being a PN.}
\end{enumerate}

Two members of our team (RC and MH) applied an alternative technique in which narrow-band CCD images taken through [S~II], 
H$\alpha$, and [O~III] filters are digitally combined. RC used medium format Pentax 6x7 camera lenses of 150-mm and 200-mm focal lengths. MH used a TMB 80-mm f/6 apochromatic refractor with a 0.8x focal reducer and the above filters.  Both observers used commercially available cooled CCD (KAF3200ME) cameras. The filtered images are assigned to the red, green, and blue channels, respectively, to produce false color images. The [O~III] emission casts a characteristic color making PN candidates easily visible (Crisp 2005).  Follow-up confirmation images were taken with the same filters on an 18" f/12.6 classical cassegrain telescope (built by RC). Candidates were then examined in the DSS as a cross check. Objects KTC1, Cr1, and HoCr1 were found this way.

It is worth noting that faint PNe have been found by visually scanning sky survey plates in the past, most notably by \cite{abell66}. More recent attempts were made by \cite{whiting02} and \cite{ellis84}.

Like all PN survey methods, the DSS approach suffers from several selection effects. One hopes that these biases differ from those of other surveys; unfortunately, this is not entirely true. For example, like other methods, this survey is biased to find high surface brightness PNe. However, unlike the on-band / off-band method, or the objective prism method, the DSS method fares better with extended objects than with compact objects, which are more easily overlooked when visually scanning the digital images. Still, a few small objects were detected (e.g., Pa~7 at 9x6 arcsec and Kn~4 at 6.6x6.2 arcsec;
see Table 2).

Figure 1 shows three excellent discoveries that are also fine examples of near-perfect circular shells. Pa~9, in particular, is almost a twin to the prototypical spherical shell PN, Abell~39 (Jacoby, Ferland, \& Korista 2001). Section 5 discusses the importance of these very circular PNe.

\medskip
\medskip
\medskip

\begin{figure}[h]
\begin{center}
\includegraphics[scale=0.65, angle=0]{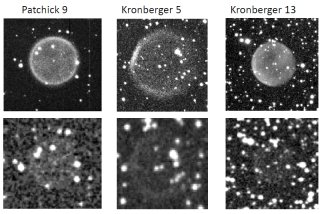}
\caption{Three circular example PNe from this survey. Pa~9 is especially nice, being a twin to the classic Abell~39. The upper panels are 5-min H$\alpha$ images taken at the WIYN 3.5-m telescope. The lower panels are the images of the candidates as seen on the POSS-II DSS (Pa~9 and Kn~13 from the red plate, while Kn~5 is from the blue). North is up and east is to the left. }\label{fig1}
\end{center}
\end{figure}

Figure 2 demonstrates two examples of highly challenging objects. Kn~25 is at the limit of detection in terms of surface brightness while KnPa~1 is extremely close to a bright double star (BD+62~582; V=7.4) and could have been considered an optical ghost by some observers. 

\begin{figure}[h]
\begin{center}
\includegraphics[scale=0.52, angle=0]{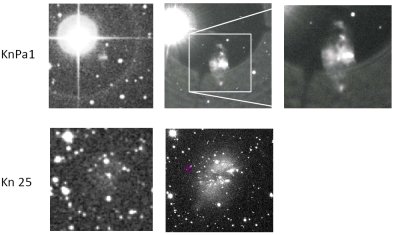}
\caption{Two of the most difficult candidates to find on the DSS. The leftmost panels are the images of the candidates as seen on the POSS-II DSS (KnPa~1 is from the red plate while Kn~25 is from the blue). The right (and center) panels are 
5-min H$\alpha$ images taken at the WIYN 3.5-m telescope. North is up and east is to the left. }\label{fig2}
\end{center}
\end{figure}


\section{Initial Results}
\subsection{Images}

The initial selection of targets contained 97 candidates, although the candidate list continues to grow with time. We have imaged 76 of these, mostly with the WIYN 3.5-m telescope and a few with the OHP 1.2-m. Fifteen candidates were rejected immediately as galaxies, apparent groupings of faint stars, and in a few cases, nothing at all. Figures 1 and 2 illustrate the effectiveness of the follow-up imaging as a critical component of the method. In addition, there is one very odd object, Sa~1 (see Figure 3), that does not match the morphological definition of a PN. It may be a supernova remnant (SNR), but spectra are needed to classify this object more definitively. 

Every imaging-based search for PNe risks the possibility of errors in classification unless high quality spectra of the objects can be obtained. The criteria for promoting a candidate to a ``true'', ``probable'', or ``possible'' PN are diverse. These include spectroscopic diagnostics (Baldwin, Phillips, \& Terlevich 1981; Kniazev, Pustilnik, \& Zucker 2008; Frew \& Parker 2009), morphology of the nebula, the presence of a low luminosity
blue central star, and, following the
discussion in Section 3, the presence and absence on various DSS bandpasses (including 2MASS). At this point in our project, spectral follow-up is scheduled for the coming year. While spectra have been obtained for about 20\% of our candidates, the depth, wavelength coverage, and resolution are generally not sufficient for a definitive classification. For consistency across our sample, we categorize our 60 candidates based on the available imaging data. That is, none of the objects is assigned to be a ``true'' PN at this stage. Those having some pathological aspect to their morphology are relegated to the list of ``possible'' PNe. The remainder, most of which exhibit a high degree of symmetry and have a blue star at the center, are assigned to the ``probable'' group.

\begin{figure}[h]
\begin{center}
\includegraphics[scale=0.50, angle=0]{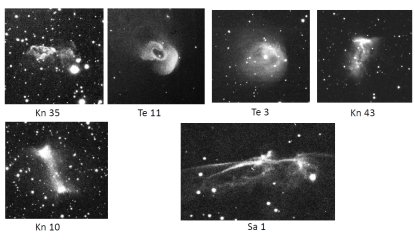}
\caption{A collection of the WIYN images for the more unusual looking objects found in this study. Perhaps Sa~1 is the oddest; it definitely does not look like a PN, but does have the filamentary characteristics of a SNR. Te~11 and Te~3 could be PNe whose shells are interacting with the local ISM. Kn~43 and Kn~10 could be the remnant tori from old bipolar PNe in which the large outer lobes have now faded into the ISM, although KN~43 is badly distorted and has a supernovae remnant-like spectrum. All images are shown with north up and east to the left. }
\label{fig3}
\end{center}
\end{figure}

Of the 60 objects remaining after the initial screening, 44 are strong candidates to be PNe. These tend to be quite regular, either round or bipolar. The final 16 objects may be PNe but fail to exhibit the nice symmetry of the first 44 targets. We defer final classifications of all 60 objects until high quality spectra of each object have been obtained. 

Figure 4 shows the 44 best PN candidates and Figure 5 shows the 16 less probable PN candidates. Tables 2 and 3 provide a summary of the PN candidates, their coordinates, and their sizes. We note that there is some unavoidable subjectivity to the classification process and several of the ``possible'' objects could have been listed in the ``probable'' group (and vice versa) if different classification criteria had been applied.

Generally, objects are given identifiers based on the first two letters of their discoverer's last name, or pairs of names in the case of a collaborative discovery. The author list of this paper provides the key to the full names of the discoverer. The exception is that Kn is used to denote Kronberger due to a conflict with another Kr (Krasnogorskaja 1962). Note that some early results from this study were presented at a conference (Jacoby et al.\ 2009) before the naming convention was properly defined, and the names in the conference paper may differ from those presented here.

\begin{figure*}[h]
\begin{center}
\includegraphics[scale=2.30, angle=0]{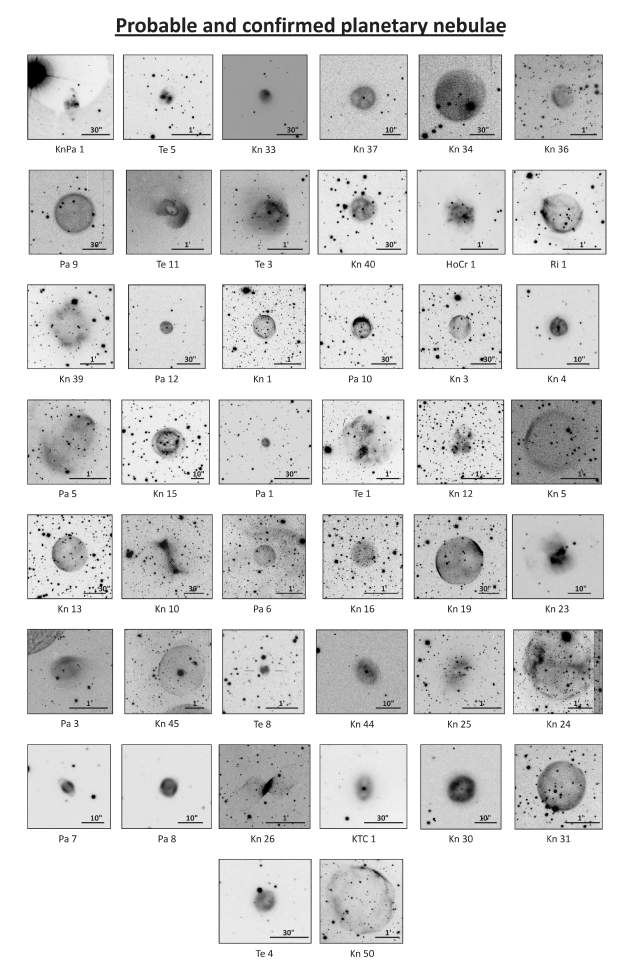}
\caption{H$\alpha$ images of the 44 best PN candidates from the 76 targets that have been imaged. Scale bars are shown for each object.
North is up and east is to the left. }\label{fig4}
\end{center}
\end{figure*}

\begin{figure*}[h]
\begin{center}
\includegraphics[scale=1.80, angle=0]{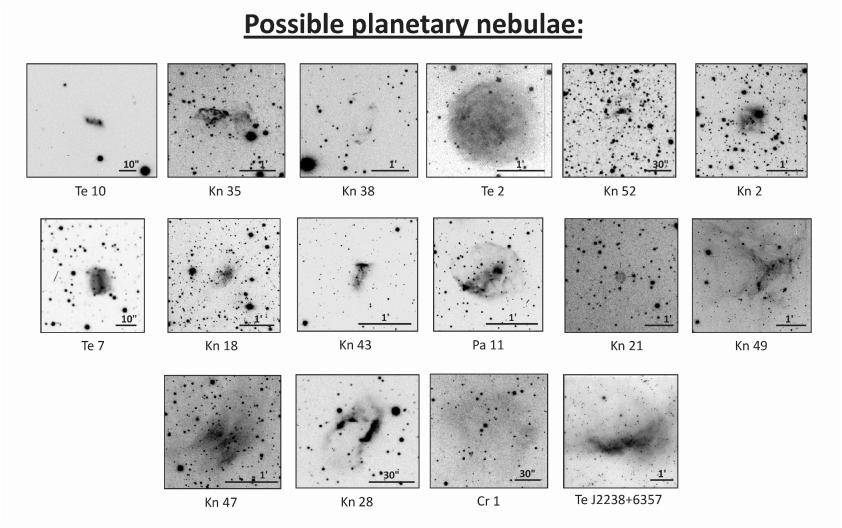}
\caption{H$\alpha$ images of the 16 possible PN candidates. Several of these may be normal PNe that have experienced an interaction with the ISM or have been distorted from symmetry by some other process.
North is up and east is to the left. }\label{fig5}
\end{center}
\end{figure*}

\subsection{Spectra}
Spectra exist for 15 of the 60 probable and possible PNe, mostly from the South African Astronomical Observatory (SAAO) 1.9-m telescope near Sutherland, and the Observatoire de Haute Provence (OHP) 1.52-m telescope 
near the village of St. Michel l'Observatoire. Candidates having some spectral information are identified in Tables 2 and 3; however, because the spectra were usually obtained for us as part of someone else's program, the diagnostic information is not always sufficient to provide a definitive classification of the object. Consequently, we plan to observe all the candidates in the upcoming year with the goal of obtaining high signal-to-noise, broad-band spectra that will allow us to classify these objects more definitively using emission-line diagnostic diagrams (Baldwin, Phillips, \& Terlevich 1981; Kniazev, Pustilnik, \& Zucker 2008).

\begin{table*}[h]
\begin{center}
\caption{Coordinates and Diameters of the Probable PNe}\label{table2}
\begin{tabular}{lccrrcl}
\\
\hline ID$^a$ & RA & Dec & l\ \ \  & b\ \ \  & Size & Comments$^b$ \\
        & J2000 & J2000 & & & arcsec & \\
\hline
\\

KnPa 1 &   03 35 57.7 & +63 16 42 &  140.2363 &  6.0767  & 36 x 17   &   \\
Te 5 &     04 03 29.6 & +52 08 25 &  150.0739 & $-$0.3084  & 35 x 25   &  DSH J0403.4+5208 \\
Kn 33 &    04 32 38.1 & +60 20 12 &  147.2155 &  8.3650  & 17 x 16   &   \\
Kn 37 &    04 44 37.8 & +37 39 15 &  165.4976 & $-$5.2624  & 31 x 28   &   \\
Kn 34 &    04 45 18.7 & +59 09 24 &  149.1758 &  8.7933  & 60 x 57   &   \\
Kn 36 &    04 55 24.5 & +52 59 15 &  154.8943 &  5.9870  & 59 x 50   &   \\
Pa 9 * &   05 37 58.0 & +17 06 18 &  189.1832 & $-$7.6966  & 53 x 53   & sp$+$  \\
Te 11 * &  05 45 58.2 & +02 21 06 &  203.1720 & $-$13.4069  & 43 x 34   & sp?, low exc  \\
Te 3 * &   06 00 34.8 & +21 41 11 &  187.9605 & $-$0.7744  & 77 x 68   & DSH J0600.5+2141; sp? \\
Kn 40 * &  06 00 47.2 & +09 28 40 &  198.6409 & $-$6.7390  & 37 x 37   &  sp? \\
HoCr 1 *&  06 21 41.0 & +23 35 13 &  188.6367 &  4.4083  & 73 x 59   &  sp$+$ \\
Ri 1 * &   06 46 24.7 & +08 29 02 &  204.8058 &  2.7496  & 70 x 62   & DSH J0646.4+0829 \\
Kn 39 &    06 59 23.8 & +18 26 49 &  197.2256 &  9.9972  & 111 x 102 &   \\
Pa 12 &    17 09 38.6 & $-$09 00 41 &   12.4614 & 17.8356  & 17 x 16   &   \\
Kn 1 &     18 35 51.5 & +10 57 19 &   40.9931 &  8.3956  & 57 x 53   &   \\
Pa 10 &    18 37 10.7 & +04 28 17 &   35.2983 &  5.2049  & 27 x 26   &   \\
Kn 3 &     18 55 21.8 & +15 11 44 &   46.9454 &  6.0161  & 29 x 28   &   \\
Kn 4 &     19 07 15.9 & +18 52 41 &   51.5351 &  5.1305  & 6.6 x 6.2 &   \\
Pa 5 &     19 19 30.6 & +44 45 44 &   76.3252 & 14.1155  & 157 x 154 &  DSH J1919.5+4445; WHI~J1919+44 \\
Kn 15$^c$ * &  19 40 40.4 & +29 30 09 &   64.5204 &  3.4068  & 30 x 23   &  DSH J1940.6+2930; sp$+$ \\
Pa 1 * &   19 47 02.7 & +29 30 26 &   65.2141 &  2.2060  & 14 x 12   &  DSH J1947.0+2930; sp$+$ \\
Te 1 * &   19 57 22.3 & +26 39 06 &   63.9272 & $-$1.2123  & 146 x 140 &  DSH J1957.3+2639; sp$+$ \\
Kn 12 &    20 03 22.5 & +21 35 52 &   60.3365 & $-$5.0288  & 54 x 47   & DSH J2003.3+2135  \\
Kn 5 &     20 03 44.3 & +12 22 26 &   52.4365 & $-$9.9044  & 47 x 47   &   \\
Kn 13 &    20 06 49.0 & +21 21 26 &   60.5575 & $-$5.8353  & 50 x 48   & DSH J2006.8+2121  \\
Kn 10 &    20 08 32.6 & +19 28 31 &   59.1704 & $-$7.1793  & 65 x 54   & DSH J2008.5+1928  \\
Pa 6 &     20 09 40.9 & +41 14 43 &   77.6494 &  4.3917  & 48 x 48   &  DSH J2009.6+4114 \\
Kn 16 &    20 20 11.5 & +24 04 38 &   64.5430 & $-$6.9406  & 48 x 47   &  DSH J2020.1+2404 \\
Kn 19 &    20 29 20.6 & +25 32 40 &   66.9485 & $-$7.8205  & 74 x 73   &  DSH J2029.3+2532 \\
Kn 23 &    20 34 26.2 & +31 18 33 &   72.3157 & $-$5.3586  & 19 x 15   &   \\
Pa 3 &     20 46 10.5 & +52 57 06 &   90.8199 &  6.1172  & 72 x 58   &  DSH J2046.1+5257 \\
Kn 45 &    20 53 03.9 & +21 00 11 &   66.5074 & $-$14.8978  & 145 x 138 &   \\
Te 8 * &   20 55 27.2 & +39 03 59 &   81.0913 & $-$3.9314  & 23 x 18   &  DSH J2055.4+3903; sp$+$ \\
Kn 44 &    21 01 20.5 & +21 40 06 &   68.2828 & $-$16.0064  & 11 x 10   &   \\
Kn 25 &    21 09 20.1 & +38 36 06 &   82.5311 & $-$6.2695  & 79 x 57   &   \\
Kn 24 &    21 13 37.7 & +37 15 38 &   82.1175 & $-$7.8014  & 190 x 190 &   \\
Pa 7$^c$ &     21 20 00.1 & +51 41 05 &   93.3034 &  1.4105  & 9 x 6     &  DSH J2120.0+5141 \\
Pa 8 &     21 20 52.8 & +50 56 40 &   92.8757 &  0.7909  & 7 x 7     &   DSH J2120.8+5056 \\
Kn 26 * &  21 23 09.3 & +38 58 13 &   84.6729 & $-$7.9641  & 110 x 51  &  Lanning~384; EWML~1; sp$+$ \\
KTC 1 &    21 28 10.8 & +58 52 36 &   99.1884 &  5.7214  & 22 x 16   &   \\
Kn 30 &    21 47 24.3 & +63 05 09 &  103.7659 &  7.2839  & 13 x 12   &   \\
Kn 31 &    22 27 39.2 & +66 44 10 &  109.4007 &  7.7201  & 80 x 80   &   \\
Te 4 &     23 27 13.2 & +65 09 23 &  114.2322 &  3.7169  & 19 x 17   &  DSH J2327.2+6509 \\
Kn 50 &    23 54 11.3 & +74 55 34 &  119.1566 & 12.4756  & 185 x 167 &   \\
\hline & & &
\end{tabular}
\end{center}
$^a$ ID entries with a ``*'' indicate that some spectroscopy data exists. \\
$^b$ An entry of ``sp$+$'' indicates that the spectral data supports classification as a PN, and ``sp?'' indicates that the spectrum neither supports nor refutes the current classification. DSH aliases are from Kronberger et al.\ (2006); WHI is from Whiting et al.\ (2007); EWML 
is from Eracleous et al.\ (2002); Lanning~384 is from Lanning \& Meakes (2000).\\
$^c$ Kn~15 (IPHASXJ194040.3+293010) and Pa~7 (IPHASXJ212000.1+514106)
were recently recovered by \cite{viironen09b}.
\end{table*}

\begin{table*}[h]
\begin{center}
\caption{Coordinates and Diameters of the Possible PNe}\label{table3}
\begin{tabular}{lccrrcl}
\\
\hline ID$^a$ & RA & Dec & l\ \ \  & b\ \ \  & Size & Comments$^b$  \\
        & J2000 & J2000 & & & arcsec & \\
\hline
\\

Te 10 & 00 13 33.6 & +67 18 04 & 119.2797  & 4.6983   & 14 x 5  &   \\
Kn 35 & 04 55 25.0 & +53 14 05 & 154.7005  & 6.1418   & 106 x 66  &   \\
Kn 38 & 05 16 16.5 & +27 27 19 & 177.6952  &$-$6.2200 & 70 x 46  &   \\
Te 2  & 05 40 44.6 & +31 44 32 & 177.0598  & 0.5795   & 122 x 117 & DSH J0540.7+3144 \\
Kn 52 & 17 08 12.4 & $-$24 29 01 & 359.0825&   9.4184 & 14 x 7  &   \\
Kn 2  & 18 32 40.0 & +13 58 02 &  43.3950  &10.4116   & 56 x 52  &   \\
Te 7$^c$ * & 18 51 47.6 & $-$00 28 30 & 32.5498&  $-$0.2958 & 14 x 12  & DSH J1851.7-0028; sp$+$  \\
Kn 18 & 19 24 06.9 & +33 52 10 &  66.7122  & 8.5756 & 64 x 27  &   \\
Kn 43 * & 19 41 07.7 & +19 08 26 &  55.5554  &$-$1.7972 & 39 x 21  &  DSH J1941.1+1908; sp$-$ \\
Pa 11 * & 19 57 59.4 & +04 47 31 &  45.0184 &$-$12.4651 & 92 x 69  &  KKH~91; WHTZ~3; sp$-$ SNR? \\
Kn 21 * & 20 41 18.0 & +27 35 06 &  70.2055 & $-$8.7815 & 29 x 25  &  DSH J2041.3+2735; sp? \\
Kn 49 & 20 55 48.0 & +65 34 00 & 101.6140 & 13.0057 & 105 x 90  &   \\
Kn 47 & 21 06 23.7 & +38 36 03 &  82.1426 & $-$5.8444 & 80 x 55  &   \\
Kn 28 & 21 22 01.0 & +55 04 30 &  95.9179 &  3.5928 & 56 x 34  & 
DSH J2122.0+5504  \\
Cr 1  & 21 49 11.1 & +57 27 25 & 100.3154 &  2.8202 &  120 x 106  &   \\
Te J2238+6357 & 22 38 11.0 &  +63 57 54  & 108.9330 &  4.7765 & 210 x 90  & Te~12  \\

\hline & 
\end{tabular}
\end{center}
$^a$ ID entries with a ``*'' indicate that some spectroscopy data exists. \\
$^b$ An entry of ``sp$-$'' indicates that the spectral data suggests that the object may not be a PN, and ``sp?'' indicates that the spectrum neither supports nor refutes the current classification. DSH aliases are from Kronberger et al.\ (2006); MPA is from Miszalski et al.\ (2008); KKH is from
Makarov et al.\ (2003); WHTZ is from Weinberger et al.\ (1999). \\
$^c$ A recent spectrum and additional imaging from the MASH-II survey (Miszalski et al.\ 2008) indicates that Te~7 (MPA~J1851-0028) is very likely a confirmed PN.
\end{table*}

Nevertheless, we do have some useful spectral data; two are shown in Figures 6 and 7. Clearly, Pa~9, which has an ideal spherical shell geometry (see Figure 1), exhibits a classic PN spectrum. Te~11, which looks distorted (see Figure 3), is probably a low excitation nebula. In projection, Te~11 is located in the neighborhood of Barnard's Loop, and perhaps its unusual morphology is related to its proximity to this star-forming region.

\begin{figure}[h]
\begin{center}
\includegraphics[scale=0.60, angle=0]{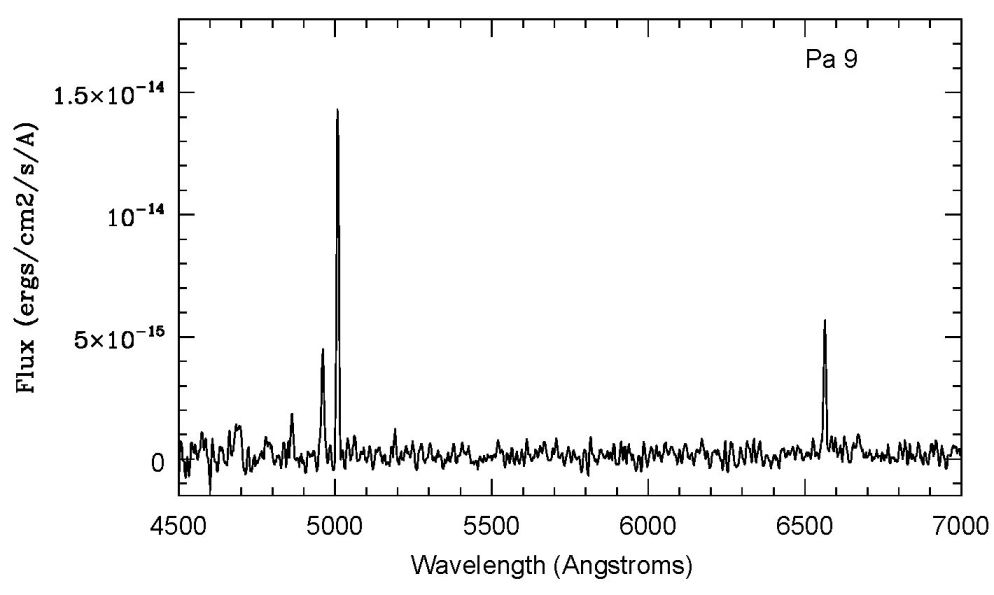}
\caption{A spectrum of Pa~9 taken at the SAAO 1.9-m. It appears to be a normal high excitation PN.}\label{fig6}
\end{center}
\end{figure}

\begin{figure}[h]
\begin{center}
\includegraphics[scale=0.60, angle=0]{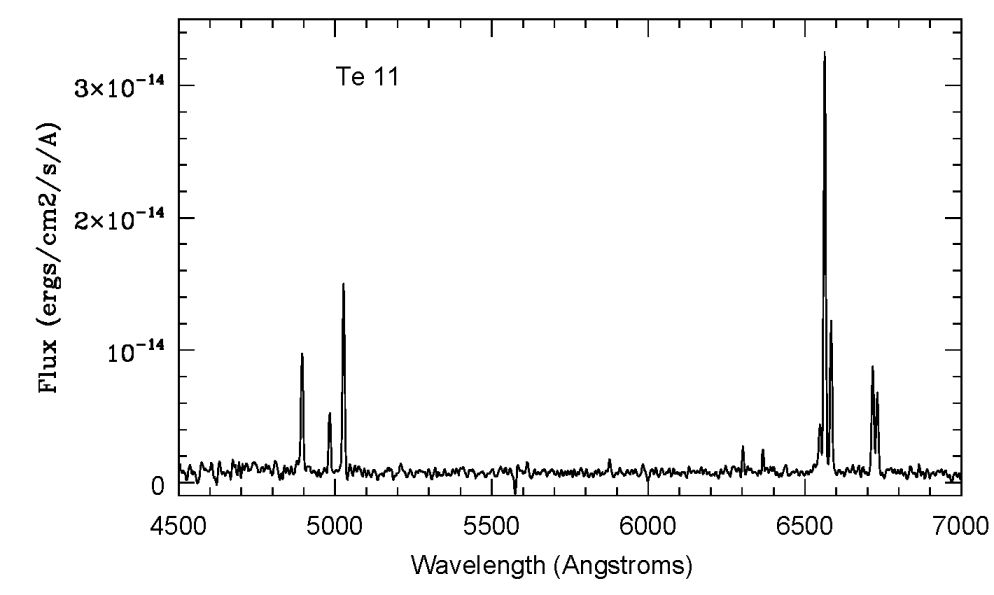}
\caption{A spectrum of Te~11 taken at the SAAO 1.9-m. This looks like a low excitation nebula, perhaps a PN. Additional information on the central star and the blue end of the spectrum would help determine the nature of this candidate more definitively.}\label{fig7}
\end{center}
\end{figure}


\section{Discussion}

\subsection{Statistics of the Survey}

This survey for faint PNe was based on the identification of 97 PN candidates found in the DSS. Of the 76 objects for which follow-up H$\alpha$ imaging is available, 60 objects are probable or possible true PNe. That is, the success rate for the method is $79\pm10$\%, although follow-up spectroscopic observations may result in a lower rate. Therefore, this study increments the Galactic census by a few percent and demonstrates that the DSS visible search method offers a useful tool for finding faint PNe with existing data sets. Because the current method is very tedious, it likely will be applied only to areas of the sky close to the Galactic plane where the return on the effort is maximized. We expect to continue the search, perhaps adding another 100--200 PNe to the overall census, and possibly many more if the search is extended. We currently have 39 candidates that have not yet been imaged.

In principle, searching the DSS data sets can be automated. Previous attempts to do so for other classes of objects such as dwarf spheroidal galaxies (Armandroff, Davies, \& Jacoby 1998) were hampered by a very high false detection rate. Instead, directed surveys like SHS and IPHAS that use the added information from narrow-band filter images have the potential to be more effective if the wide-field data can be obtained.
Although our group does not plan to develop the software for automation, the all-sky nature of the existing data sets may be alluring enough for groups with greater resources to initiate a PN search of this kind. Probably, the most significant result from that kind of survey would be the identification of numerous halo PNe.

\subsection{Spherical PNe}

For the moment, we assume that all 60 objects shown in Figures 4 and 5
are true PNe. Of these, $\sim$20\% are very round, consistent with the
stringent definition for ``round'' adopted by \cite{soker97}. That is,
a round PN has no departure from sphericity, such as internal
structures (knots, arcs), and is not a pole-on elliptical. These
fractions are very similar to the statistics seen in the larger samples
of PNe found by \cite{parker06} and \cite{viironen09a}. 

It is possible that the large fraction of round objects is a selection effect. A round candidate is more likely to be included in the sample, especially in the ``probable'' group of Figure 4, than some less symmetrical form. It may also be true that a round morphology is more likely to be identified than other shapes during the initial screening of the DSS images. The considerable morphological diversity of the PN candidates in Figures 3--5, however, argues that this effect is not a major factor. If a selection effect is present, it is difficult to quantify and its importance would be highly dependent on the observer. For this paper, we assume that selection effects are minor.

The large number of round PNe found in the recent faint surveys is in strong contrast to the more familiar collection of PNe where the fraction of known round PNe is between $\sim$4\% (Soker 1997) and 10\% (Soker 2002). As a class, the near-perfect, ideal geometry, spherical PNe seem to be drawn exclusively from a population of PNe having low surface brightness (Jacoby et al.\ 2001; Pierce et al.\ 2004).

Thus, we have an intriguing quandary. Where do low surface brightness round PNe come from if not from high surface brightness round PNe, for which there are few, if any, examples? \cite{soker02} argues that round PNe have a different formation history than non-round PNe due to the absence of a companion star or planet. In fact, \cite{soker05} predict that there has been a hidden population of exactly these kinds of PNe waiting to be found, and that $\sim$30\% of all PNe fall in this class. The underlying basis for their models is that deviations from sphericity are due to a superwind (Kwok, Purton, \& Fitzgerald 1978) that has been made non-spherical due to spin-up of the AGB star by a companion, possibly a planet (Soker 2002). Soker's interpretation, then, is that spherical PNe are the descendants of stars that do not have a nearby companion (e.g., a low mass star, substellar object, or a planet) and a significant fraction of the brighter, more familiar PNe are descendents of a process involving a nearby object. This view, though, fails to explain why the round PNe tend to have a higher average scale height above the plane than the non-round PNe (Frew 2008).

Also, we do not see a large fraction of binary stars among the central star population of PNe. \cite{miszalski09} find this fraction to be 10-20\%. The jury is still out, though, on the true fraction of binary central stars because small companions are extremely difficult to detect; that is, we don't know the true fraction of PNe central stars that have been influenced by small companions.


\section{Conclusions}

We have demonstrated a method for finding dozens to hundreds of faint PNe using digital sky survey data. We do not expect to find the missing thousands, or tens of thousands of PNe that are predicted to be hiding in the Galaxy, obstructed by dust. If those objects truly exist, other techniques are needed. Br$\gamma$ is a good choice for a future emission line survey because it becomes one of the brightest lines from PNe when the visual extinction exceeds $\sim$12 magnitudes (Jacoby \& Van de Steene 2004). With the recent development of wide-field IR imagers, it is time to begin a survey of this type.

Still, the optical surveys continue to be useful because they probe more deeply than earlier PN surveys. An intriguing result is the emergence of a large number of very spherical PNe at low surface brightness in this and other recent surveys. Their origin and significance remains an active topic for discussion.

\section*{Acknowledgments} 

We wish to thank Drs. Kimberly Herrmann and Robin Ciardullo for allowing us to obtain images for several of these PN candidates during their observing time, and for collecting some of the images for us. We thank the anonymous referee for providing many suggestions to improve the presentation of this paper.


\begin{thebibliography}{}

\bibitem[Abell (1966)]{abell66} Abell, G.O. 1966, ApJ, 144, 259
\bibitem[Acker et al.\ (1992a)]{acker92a} Acker, A., Cuisinier, F., Stenholm, B., \& Terzan, A. 1992, A\&A, 264, 217
\bibitem[Acker et al.\ (1992b)]{acker92b} Acker, A., Marcout, J., Ochsenbein, F., Stenholm, B., Tylenda, R., \& Schohn, C. 1992, The Strasbourg-ESO Catalogue of Galactic Planetary Nebulae., (ESO-Garching, Germany)
\bibitem[Acker et al.\ (1996)]{acker96} Acker, A., Marcout, J., Ochsenbein, F., with the collaboration of S. Beaulieu \& G. Jacoby 1996, First Supplement to the Strasbourg-ESO Catalogue of Galactic Planetary Nebulae, Publication de l'Observatoire de Strasbourg
\bibitem[Armandroff, Davies, \& Jacoby(1998)]{armandroff98} Armandroff, T.E., Davies, J.E., \& Jacoby, G.H. 1998, AJ, 116, 2287
\bibitem[Baldwin, Phillips, \& Terlevich (1981)]{baldwin81} Baldwin, J.A., Phillips, M.M., \& Terlevich, R. 1981, PASP, 93, 5 
\bibitem[Cahn \& Wyatt (1976)] {cahn76} Cahn, J.H., \& Wyatt, S.P. 1976, ApJ, 210, 508
\bibitem[Cappellaro et al.\ (1990)]{cappellaro90} Cappellaro, E., Turatto, M., Salvadori, L., \& Sabbadin, F. 1990, A\&AS 86, 503
\bibitem[Ciardullo et al.\ (1989)]{ciardullo89} Ciardullo, R.B., et al.\ 1989, ApJ, 339, 53
\bibitem[Ciardullo et al.\ (2005)]{ciardullo05} Ciardullo, R., Sigurdsson, S., Feldmeier, J., \& Jacoby, G.H. 2005, ApJ, 629, 499
\bibitem[Crisp (2005)]{crisp05} Crisp, R., 2005, S\&T, 110, \#2, 112
\bibitem[De Marco \& Moe (2005)]{demarco05} De Marco, O., \& Moe, M. 2005, AIP, 804: Planetary Nebulae as Astronomical Tools, (New York: AIP), 169
\bibitem[Drew et al.\ (2005)]{drew05} Drew, J.E., et al.\ 2005, MNRAS, 362, 753
\bibitem[Ellis, Grayson, \& Bond (1984)]{ellis84} Ellis, G.L. , Grayson, E.T., \& Bond, H.E. 1984, PASP, 96, 283
\bibitem[Frew (2008)]{frew08} Frew, D.J. 2008, PhD thesis, Macquarie University
\bibitem[Frew \& Parker (2009)]{frew09} Frew, D.J., \& Parker, Q.A. 2009, this proceedings, in press
\bibitem[Howell et al.\  (2003)]{howell03} Howell, S.B., Everett, M.E., Tonry, J.L., Pickles, A, \& Dain, C. 2003, PASP, 115, 1340
\bibitem[Ishida \& Weinberger (1987)]{ishida87} Ishida, K, \& Weinberger, R. 1987, A\&A, 178, 221
\bibitem[Jacoby (1980)]{jacoby80} Jacoby, G.H. 1980, ApJS, 42, 1
\bibitem[Jacoby, Ferland, \& Korista (2001)]{jacoby01} Jacoby, G.H., Ferland, G.J., \& Korista, K.T. 2001, ApJ, 560, 272
\bibitem[Jacoby \& Van de Steene (2004)]{jacoby04} Jacoby, G.H., \& Van de Steene, G. 2004, A\&A, 419, 563
\bibitem[Jacoby et al.\ (2009)]{jacoby09} Jacoby, G.H., et al.\ 2009, in Asymmetrical Planetary Nebulae IV, eds. R. Corradi, A. Manchado, \& N. Soker, 199
\bibitem[Kniazev, Pustilnik, \& Zucker, D. (2008)]{kniazev08} Kniazev, A.Y., Pustilnik, S.A., \& Zucker, D.B. 2008, MNRAS, 384, 1045
\bibitem[Kohoutek (2001)]{kohoutek01} Kohoutek, L. 2001, A\&A, 378, 843
\bibitem[Krasnogorskaja (1962)]{kras62} Krasnogorskaja, A. 1962, ATsir, 230, 11
\bibitem[Kronberger et al.\ (2006)]{kronberger06} Kronberger, M. et al.\ 2006, A\&A, 447, 921
\bibitem[Lanning \& Meakes (2000)] {lanning00} Lanning, H.H., \& Meakes, M. 2000, PASP, 112, 251
\bibitem[Makarov, Karachentsev, \& Burenkov (2003)]{makarov03} Makarov, D.I., Karachentsev, I.D., \& Burenkov, A.N. 2003, A\&A, 405, 951
\bibitem[Miszalski et al.\ (2008)]{miszalski08} Miszalski, B., Parker, Q.A., Acker, A., Birkby, J.L., Frew, D.J., \& Kovacevic, A. 2008, MNRAS, 384, 525
\bibitem[Miszalski et al.\ (2009)]{miszalski09} Miszalski, B., Acker, A., Moffat, A.F.J., Parker, Q.A., \& Udalski, A., 2009, A\&A, 496, 813
\bibitem[Kwok, Purton, \& Fitzgerald (1978)]{kwok78} Kwok, S., Purton, C.R., \& Fitzgerald, P.M. 1978, ApJ, 219, L125
\bibitem[Moe \& De Marco (2006)]{moe06} Moe, M., \& De Marco, O. 2006, ApJ, 650, 916
\bibitem[Napiwotzki (2001)]{napiwotzki01} Napiwotzki, R. 2001, A\&A, 367, 973
\bibitem[Parker et al.\ (2005)]{parker05} Parker, Q.A., et al.\ 2005, MNRAS 362, 689
\bibitem[Parker et al.\ (2006)]{parker06} Parker, Q.A., et al.\ 2006,
MNRAS, 373, 79
\bibitem[Peimbert (1990)]{peimbert90} Peimbert, M. 1990, RevMexAA, 20, 119
\bibitem[Peyaud (2005)]{peyaud05} Peyaud, A. 2005, PhD thesis, Observatoire de Strasbourg
\bibitem[Phillips (2002)]{phillips02} Phillips, J. 2002, ApJS, 139, 199
\bibitem[Pierce et al.\  (2004)]{pierce04} Pierce, M.J., Frew, D.J., Parker, Q.A., \& K\"oppen, J. 2004, 
PASA, 21, 334
\bibitem[Reid \& Parker (2006)]{reid06} Reid, W.A., \& Parker, Q.A. 2006, MNRAS, 373, 521
\bibitem[Reid et al.\ (2009)]{reid09} Reid, M.J. et al.\ 2009, arXiv0902.3913
\bibitem[Soker (1997)]{soker97} Soker, N. 1997, ApJS, 112, 487
\bibitem[Soker (2002)]{soker02} Soker, N. 2002, A\&A, 386, 885
\bibitem[Soker \& Subag (2005)]{soker05} Soker, N., \& Subag, E. 2005, AJ 130, 2717
\bibitem[Tovmassian et al.\ (2007)]{tovmassian07} Tovmassian, G., et al.\ 2007, astro-ph/0709.4016
\bibitem[Viironen et al.\ (2009a)]{viironen09a} Viironen, K. et al.\ 2009a, in Asymmetrical Planetary Nebulae IV, eds. R. Corradi, A. 
Manchado, \& N. Soker, 79
\bibitem[Viironen et al.\ (2009b)]{viironen09b} Viironen, K., et al.\
2009b, arXiv0906.1792
\bibitem[Wareing et al.\ (2006)]{wareing06} Wareing, C.J., et al.\ 2006,
MNRAS, 366, 387
\bibitem[Weinberger et al.\ (1999)]{weinberger99} Weinberger, R., Hartl, H., Temporin, S., \& Zanin, C. 1999, New Perspectives on the Interstellar Medium, ASPC, 168, 142
\bibitem[Whiting, Hau, \& Irwin (2002)]{whiting02} Whiting, A.B.,, Hau, G.K.T., \& Irwin, M. 2002, ApJS, 141, 123
\bibitem[Whiting, Hau, Irwin, \& Verdugo (2007)]{whiting07} Whiting, A.B.,, Hau, G.K.T., Irwin, M., \& Verdugo, M. 2007, AJ, 133, 715

\end{thebibliography}
\end{document}